\documentclass[useAMS]{mn2e}
\usepackage{epsfig,times,natbib_vw}

\bibpunct[, ]{(}{)}{;}{a}{}{,}

\def \aj {AJ}
\def \mnras {MNRAS}
\def \apj {ApJ}

\def \aap {A\&A}

\newcommand{\apjvec}[1]{\mbox{\boldmath{$#1$}}}
\newcommand{\apjmat}[1]{{\mathbf{#1}}}

\title[Reliable eigenspectra]{Reliable Eigenspectra for New Generation Surveys}
\author[Budav\'ari, Wild, et~al.]{
\parbox[t]{\textwidth}
{\raggedright Tam\'as Budav\'ari$^{1}$\thanks{E-mail: budavari@jhu.edu}, Vivienne Wild$^2$, Alexander S. Szalay$^{1,2}$, L\'aszl\'o Dobos$^3$ \\
and Ching-Wa Yip$^1$}
\vspace*{6pt}\\
$^1$Department of Physics and Astronomy, The Johns Hopkins
University, 3701 San Martin Drive, Baltimore, MD 21218, USA \\
$^2$Max Planck Institute for Astrophysics, Karl-Schwarzschild Str.~1,
85741 Garching, Germany \\
$^3$E\"otv\"os Lor\'and University, Department of Physics of Complex Systems, P\'azm\'any P.\ s\'et\'any 1/A, Budapest, 1117, Hungary
}

\begin{document}

\maketitle

\begin{abstract}
We present a novel technique to overcome the limitations of the
applicability of Principal Component Analysis to typical real-life
data sets, especially astronomical spectra. Our new approach addresses
the issues of outliers, missing information, large number of
dimensions and the vast amount of data by combining elements of robust
statistics and recursive algorithms that provide improved eigensystem
estimates step-by-step.  We develop a generic mechanism for deriving
reliable eigenspectra without manual data censoring, while utilising
all the information contained in the observations.
We demonstrate the power of the methodology on the attractive
collection of the VIMOS VLT Deep Survey spectra that manifest most of
the challenges today, and highlight the improvements over previous
workarounds, as well as the scalability of our approach to collections
with sizes of the Sloan Digital Sky Survey and beyond.
\end{abstract}


\section{Motivation}

As modern telescopes collect more and more data in our exponential world, where the
size of the detectors essentially follows Moore's law, the kind of statistical
challenges astronomers face in analysing the observations change dramatically in nature.
We need algorithms that scale well in time and complexity with the
volume of the data, while obeying the constraints of today's
computers.  But the large data volume is only one of the consequences
of this trend.  With more observations in hand, the number of
problematic detections also increases.
In addition to the elegant mathematical theorems that work miracles on
textbook examples, scientists need to develop methodologies that provide
reliable results that are robust in the statistical sense when applied to
real-life data.

One particular multi-variate analysis technique, which is widely
accepted and popular not only in astronomy but also in genetics,
imaging and many other fields, is Principal Component Analysis
(PCA). For its simple geometric meaning and straightforward
implementation via Singular Value Decomposition (SVD), it has been
utilised in many areas of the field including classification of
galaxies and quasars \citep{francis92,connolly95,connolly99,yip04a,yip04b},
spectroscopic and photometric redshift estimation
\citep{glazebrook98,budavari99,budavari00}, sky subtraction
\citep{wild05}, and highly efficient optical spectral indicators
\citep{wild07}.
However, direct application of the classic PCA  to real
data is almost always impossible; the reasons are usually three-fold:
(1) The technique is extremely sensitive to outliers.
With smaller datasets, scientists would ``clean up'' the sample by completely
removing the ``obvious'' outliers in one or more projections and analyse the
remaining subset. The problem with this approach is that it is
subjective, and it becomes impractical for large datasets.
(2) Another problem is missing measurements in the data vectors, e.g.,
pixels of a spectrum. There are various reasons for this to occur:
strong night sky lines, cosmic rays or simply because of the redshift
that yields different restframe wavelength coverage for the
spectra. The application of PCA implies the assumption of a Euclidean
metric, but it is not clear how to calculate Euclidean distances when
data is missing from our observed vectors.
(3) Last but not least, the memory requirement of SVD is significant
as the entire dataset is stored in memory while the decomposition is
computed. For example, the Sloan Digital Sky Survey (SDSS) Data
Release 6 contains a million spectra with 4000 resolution
elements each. While machines certainly exist today that have the
required amount of RAM ($\sim\!50$GB), typical workstations have
lesser resources.  Additionally, in most situations, we only seek a
small number of eigenvectors associated with the largest eigenvalues,
and SVD computes all the singular vectors in vain.

The optical spectra of the VIMOS VLT Deep Survey \citep[VVDS;][]{vvds}
provides an extremely attractive dataset for galaxy evolution studies
at high redshift, yet, due to their generally low signal-to-noise
ratio they are unsuitable for traditional PCA.  In this case, the
challenge does not lie in the volume of the data set, rather in the
natural limitations of high redshift observations. Thanks to the
careful processing by the VVDS Team, the spectra are well calibrated
and each one contains valuable information for a PCA analysis.

Our goal is to develop an algorithm to address all of the above issues in a way that
is true to the spirit of the PCA and maintains its geometrically meaningful
properties.
In \S{}\ref{sec:pca} we detail the various layers of our solution to
the problem, and in \S{}\ref{sec:vvds} we compare the performance of
different techniques when applied to the VVDS spectra.  In
\S{}\ref{sec:disc} we evaluate the results of the methods and analyse
the emerging physical features.  \S{}\ref{sec:sum} concludes our
study.

\section{Streaming PCA} \label{sec:pca}

Our approach to the analysis is not the classical one. Instead of
working with a {\it data set}, we aim to formulate the problem using
the concept of a {\it data stream}. We want to incrementally improve
our understanding of the properties of the data, deriving better and
better eigenspectra through the incremental addition of new
observations.

We develop an algorithm to recursively calculate the quantities of interest.
As the first step and an illustration of the concept, we look at the calculation
of the sample mean,
\begin{equation}
\apjvec{\mu} = \frac{1}{N} \sum_{n=1}^N \apjvec{x}_n
\end{equation}
where $\left\{\apjvec{x}_n\right\}$ are the $N$ observation vectors. One can define
a series of estimates as
\begin{equation}
\apjvec{\mu}_n = \frac{n-1}{n} \apjvec{\mu}_{n-1} + \frac{1}{n} \apjvec{x}_n
\end{equation}
It is easy to see that $\apjvec{\mu}_1 = \apjvec{x}_1$ and
$\apjvec{\mu}_N = \apjvec{\mu}$. This iterative formula
is the key to our new procedure, where the best estimate of
the sample mean at each step is
\begin{eqnarray}
\apjvec{\mu} & = & \gamma \apjvec{\mu}_{\rm{prev}} + (1-\gamma) \apjvec{x} \\
             & = & \apjvec{\mu}_{\rm{prev}} + (1-\gamma) \apjvec{y}
\end{eqnarray}
where we introduced the centered variable $\apjvec{y} = \apjvec{x} -
\apjvec{\mu}_{\rm{prev}}$. The weight parameter $\gamma$ lies between
0 and 1, and may be a function of both the observation vector and the
iteration step.

\subsection{Updating the Eigensystem}

The calculation of the sample covariance matrix is essentially
identical to that of the mean, except we average the outer products of
the vectors.
Let us solve for the eigenspectra that belong to the largest $p$
eigenvalues that account for most of the sample variance. This means
that the $\apjmat{E}_p\apjmat{\Lambda}_p\apjmat{E}_p^T$ is a good
approximation to the full covariance matrix, where
$\left\{\apjmat{\Lambda}_p,\apjmat{E}_p\right\}$ is the truncated
eigensystem.  Hence, the recursion takes the form of
\begin{eqnarray}
\apjmat{C} & = & \gamma \apjmat{C}_{\rm{prev}} + (1-\gamma) \apjvec{y} \apjvec{y}^T \\
             & \approx &  \gamma \apjmat{E}_p \apjmat{\Lambda}_p \apjmat{E}_p^T + (1-\gamma) \apjvec{y} \apjvec{y}^T
\end{eqnarray}
Following \citet{li03}, we write the covariance matrix as the
product of some matrix $\apjmat{A}$ and its transpose
\begin{equation}
\apjmat{C} \approx \apjmat{A} \apjmat{A}^T
\end{equation}
where the matrix $\apjmat{A}$ has only $(p+1)$ columns, and is thus
much smaller than the covariance matrix. The columns of $\apjmat{A}$
are the constructed from the previous eigenvalues $\lambda_k$ and
eigenspectra $\apjvec{e}_k$, as well as the new observation vector
$\apjvec{y}$,
\begin{eqnarray}
\apjvec{a}_{k} & = &  \apjvec{e}_{k} \sqrt{\gamma \lambda_k},\ \ \ \ \    k=1\dots{}p\\
\apjvec{a}_{p+1} & = & \apjvec{y} \sqrt{1 - \gamma}
\end{eqnarray}
If $\apjmat{A} = \apjmat{U} \apjmat{W} \apjmat{V}^T$ then the
eigensystem of the covariance $\apjmat{C}$ will have eigenvalues of
$\apjmat{\Lambda} = \apjmat{W}^2$ and eigenspectra equal to the
singular-vectors, $\apjmat{E} = \apjmat{U}$. Therefore, this formalism
allows us to update the eigensystem by solving the SVD of the much
smaller $\apjmat{A}$ leading to a significant decrease in
computational time.

Following the above procedure, one can update the truncated
eigensystem step-by-step by adding the observed spectra one-by-one to
build the final basis.  A natural starting point for the iteration is
to run SVD on a small subset of observation vectors first and proceed
with the above updates from there.

\subsection{Robustness against Outliers}

Before we turn to making the algorithm robust, to understand the
limitations of PCA let us first review the geometric problem that PCA
solves. The classic procedure essentially fits a hyperplane to the
data, where the eigenspectra define the projection matrix onto this
plane. If the truncated eigensystem consists of $p$ eigenspectra in
the matrix $\apjmat{E}_p$, the projector is $\apjmat{E}_p
\apjmat{E}_p^T$, and the residual of the fit for the $n$th observation
vector is written as
\begin{equation}
\apjvec{r}_n = (\apjmat{I} - \apjmat{E}_p \apjmat{E}_p^T) \apjvec{y}_n
\end{equation}
Using this notation, PCA simply solves the minimization problem
\begin{equation}
\min \frac{1}{N} \sum_{n=1}^N r_n^2
\end{equation}
where $r_n \equiv |\apjvec{r}_n|$. The sensitivity of PCA to outliers
comes from the fact that the sum will be dominated by the extreme
values in the data set.

Over the last couple of decades, a number of improvements have been
proposed to overcome this issue within the framework of robust
statistics \citep[e.g., see][for a concise overview]{maronna06}.
The current state-of-the-art technique introduced by \citet{maronna05} is
based on a robust M-estimate \citep{huber} of the scale, called M-scale.
Here we solve the new minimisation problem
\begin{equation}
\min \sigma^2
\end{equation}
where $\sigma^2$ is an M-scale of the residuals $r^2$, which
satisfies the equation
\begin{equation} \label{eq:mscale}
\frac{1}{N}\sum_{n=1}^N \rho\left(\frac{r_n^2}{\sigma^2}\right) = \delta
\end{equation}
where $\rho$ is the robust function.
Usually a robust $\rho$-function is bound and assumed to be scaled to values
between \mbox{$\rho(0)\!=\!0$} and \mbox{$\rho(\infty)\!=\!1$}.
The parameter $\delta$ controls the breakdown point where the estimate
explodes due to too much outlier contamination.
It is straightforward to verify that in the non-robust maximum likelihood
estimation (MLE) case with $\rho(t)=t$ and $\delta=1$,
we recover the classic minimization problem with $\sigma$ being the
root mean square (RMS).

By implicit differentiation the robust solution yields a very
intuitive result: the mean is a weighted average of the observation
vectors, and the hyperplane is derived from the eigensystem of a
weighted covariance matrix,
\begin{eqnarray}
\apjvec{\mu} & = & \left(\sum w_n \apjvec{x}_n\right) \Big/ \left(\sum w_n\right) \\
\apjmat{C} & = & \sigma^2 \left(\sum w_n (\apjvec{x}_n\!\!-\!\!\apjvec{\mu})(\apjvec{x}_n\!\!-\!\!\apjvec{\mu})^T\right)\Big/\left(\sum w_n r_n^2\right)
\end{eqnarray}
where $w_n = W(r_n^2/\sigma^2)$ and $W(t)=\rho'(t)$.
The weight for each observation
vector depends on $\sigma^2$, which suggests the
appropriateness of an iterative solution, where in every
step we solve for the eigenspectra and use them to calculate
a new $\sigma^2$ scale; see \citet{maronna05} for details.
One way to obtain the solution of eq.(\ref{eq:mscale}) is to
re-write it in the intuitive form of
\begin{equation}
\sigma^2 = \frac{1}{N\delta} \sum_{n=1}^N w^{\star}_n r_n^2
\end{equation}
where the weights are $w^{\star}_n = W^{\star}(r_n^2/\sigma^2)$ with
\mbox{$W^{\star}(t)=\rho(t)/t$}.
Although, this is not the solution as the right hand side contains $\sigma^2$ itself,
it can be shown that its iterative re-evaluation converges to the solution.

We take this approach one step further. By recursively calculating the eigenspectra instead
of the classic method,
we can allow for a simultaneous solution for the scale $\sigma^2$, as well.
The recursion equation for the mean is formally almost identical to
the classic case, and we introduce new equations to propagate the
weighted covariance matrix and the scale,
\begin{eqnarray}
\apjvec{\mu} & = & \gamma_1 \apjvec{\mu}_{\rm{prev}} + (1-\gamma_1) \apjvec{x} \\
\apjmat{C} & = & \gamma_2 \apjmat{C}_{\rm{prev}} + (1-\gamma_2)\ \sigma^2 \apjvec{y}\apjvec{y}^T\!\!/r^2 \\
\sigma^2  & =  & \gamma_3 \sigma^2_{\rm{prev}} + (1-\gamma_3)\ w^{\star} r^2 /\delta \label{eqn:sig2}
\end{eqnarray}
where the $\gamma$ coefficients depend on the running sums of $1$, $w$ and $wr^2$
denoted below by $u$, $v$ and $q$, respectively.
\begin{eqnarray}
\gamma_1 \ = & \alpha v_{\rm{prev}} / v \ & {\rm{with}} \ \ \ \ v = \alpha v_{\rm{prev}} + w \\
\gamma_2 \ = & \alpha q_{\rm{prev}} / q \ & {\rm{with}} \ \ \ \ q = \alpha q_{\rm{prev}} + w r^2 \\
\gamma_3 \ = & \alpha u_{\rm{prev}} / u \ & {\rm{with}} \ \ \ \ u = \alpha u_{\rm{prev}} + 1
\end{eqnarray}
%
The parameter $\alpha$ introduced here,
which takes values between 0 and 1, adjusts the
rate at which the evolving solution of the eigenproblem {\it forgets}
about past observations. It sets the characteristic width of the
sliding window over the stream of data; in other words, the effective
sample size.%
\footnote{For example, the sequence $u$ rapidly converges to $1/(1-\alpha)$.}
The value $\alpha=1$ corresponds to the classic case of infinite memory.
Since our iteration starts from a non-robust set of eigenspectra, a
procedure with $\alpha < 1$ is able to eliminate the effect of the initial
transients.
Due to the finite memory of the recursion, it is clearly
disadvantagous to put the spectra on the stream in a systematic order;
instead they should be randomized for best results.

It is worth noting that robust ``eigenvalues'' can be computed for any
eigenspectra in a consistent way, which enables a meaningful comparison of
the performance of various bases.
To derive a robust measure of the scatter of the data along a given
eigenspectrum $\apjvec{e}$, one can project the data on it, and
formally solve the same equation as in eq.(\ref{eq:mscale}) but
with the residuals replaced with the projected values, i.e., for the
$k$th eigenspectrum $r_n = \apjvec{e}_k \apjvec{y}_n$. The resulting
$\sigma^2$ value is a robust estimate of $\lambda_k$.

\subsection{Missing Entries in Observations}

The other common challenge is the presence of gaps in the
observations, i.e., missing entries in the data vectors.  Gaps
emerge for numerous reasons in real-life measurements. Some cause the
loss of random snippets while others correlate with physical
properties of the sources. An example of the latter is the wavelength
coverage of objects at different redshifts: the observed wavelength
range being fixed, the detector looks at different parts of the
electromagnetic spectrum for different extragalactic objects.

Now we face two separate problems.
Using PCA implies that we believe the Euclidean metric to be
a sensible choice for our data, i.e., it is a good measure of similarity.
Often one needs to normalize the observations
so that this assumption would hold.
For example, if one spectrum is identical to another but the source is brighter
and/or closer,
their distance would be large. The normalisation guarantees that
they are close in the Euclidean metric.
So firstly, we must normalise every spectrum
before it is entered into the streaming algorithm. This step is difficult
to do in the presence of incomplete data, hence we also have to
``patch'' the missing data.

Inspired by \citet{gappykl}, \citet{connolly99} proposed a solution,
where the gaps are filled by an unbiased reconstruction using a
pre-calculated eigenbasis. A final eigenbasis may be calculated
iteratively by continuously filling the gaps with the previous
eigenbasis until convergence is reached \citep{yip04a}. While
\citet{connolly99} allowed for a bias in rotation only, the method has
recently been extended to accommodate a shift in the normalisation of
the data vectors \citep[][Lemson et~al., in preparation]{wild07}.
Of course, the new algorithm
presented in this paper can use the previous eigenbasis to fill gaps
in each input data vector as they are input, thus avoiding the need
for multiple iterations through the entire dataset.

The other problem is a consequence of the above solution. Having
patched the incomplete data by the current best understanding of the
manifold, we have artificially removed the residuals in the bins of
the missing entries, thus decreased the length of the
residual vector. This would result in increasingly large weights
being assigned to spectra with the largest number of empty pixels.
One solution is to calculate the residual vector using higher-order eigenspectra.
The idea is to
solve for not only the first $p$ eigenspectra but
a larger $(p\!+\!q)$ number of components and estimate the residuals in the missing
bins using the
difference of the reconstructions on the two different truncated bases.

\section{The VVDS Spectra} \label{sec:vvds}

\begin{figure*}
\includegraphics[scale=0.7]{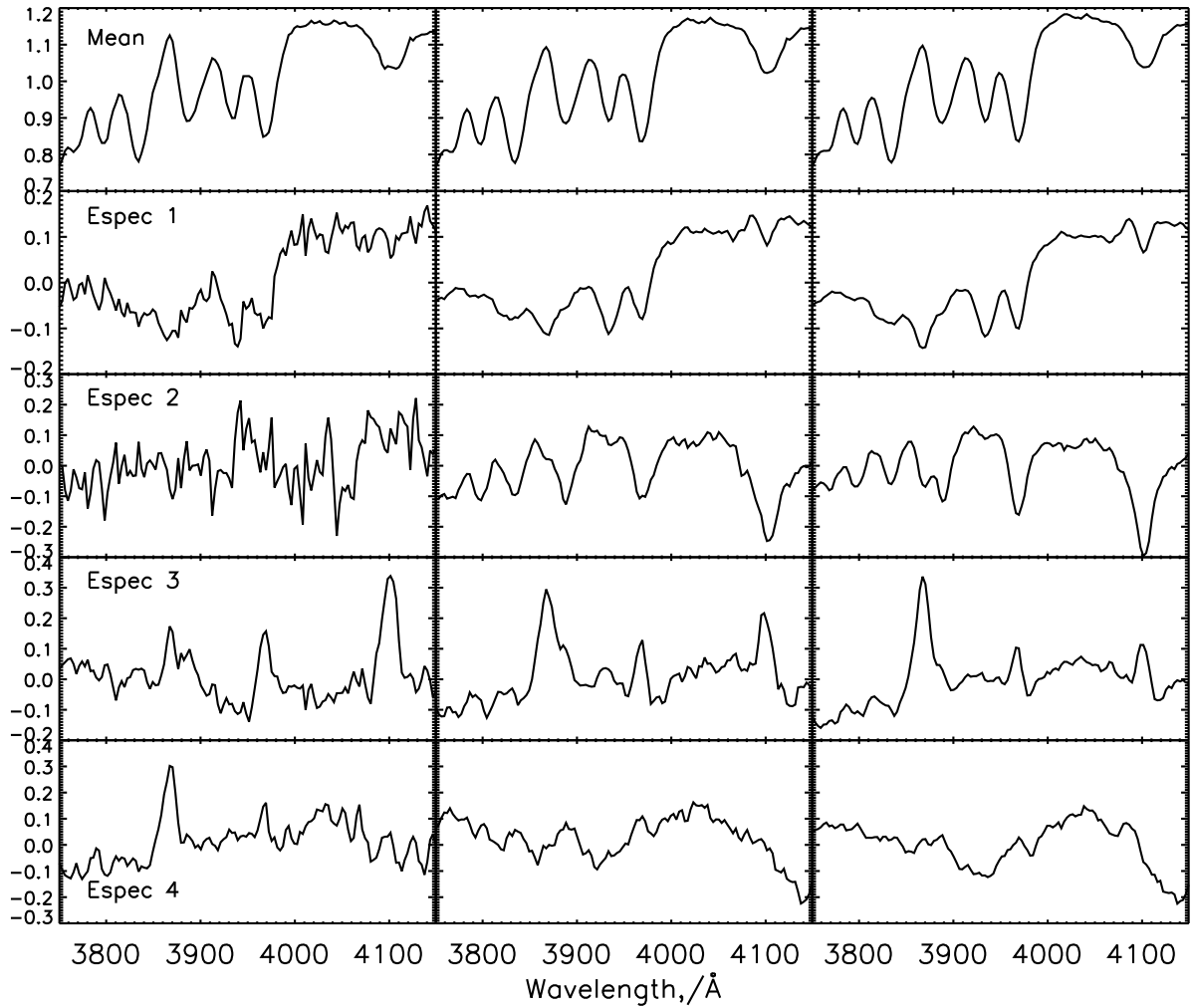}
\caption{The mean spectrum and top four eigenspectra for the VVDS
galaxies. The eigenspectra have been inverted where necessary to make
the physical features easier to identify (i.e. absorption lines in
absorption and emission lines in emission) {\it Left:} The result from
classic PCA on 3485 spectra. {\it Center:} The result from classic PCA
with iterative removal of outliers. The final dataset contains 2675
spectra. {\it Right:} The result from the new iterative-robust PCA
algorithm.  }\label{fig:espec}
\end{figure*}

\begin{figure*}
\includegraphics[scale=0.5]{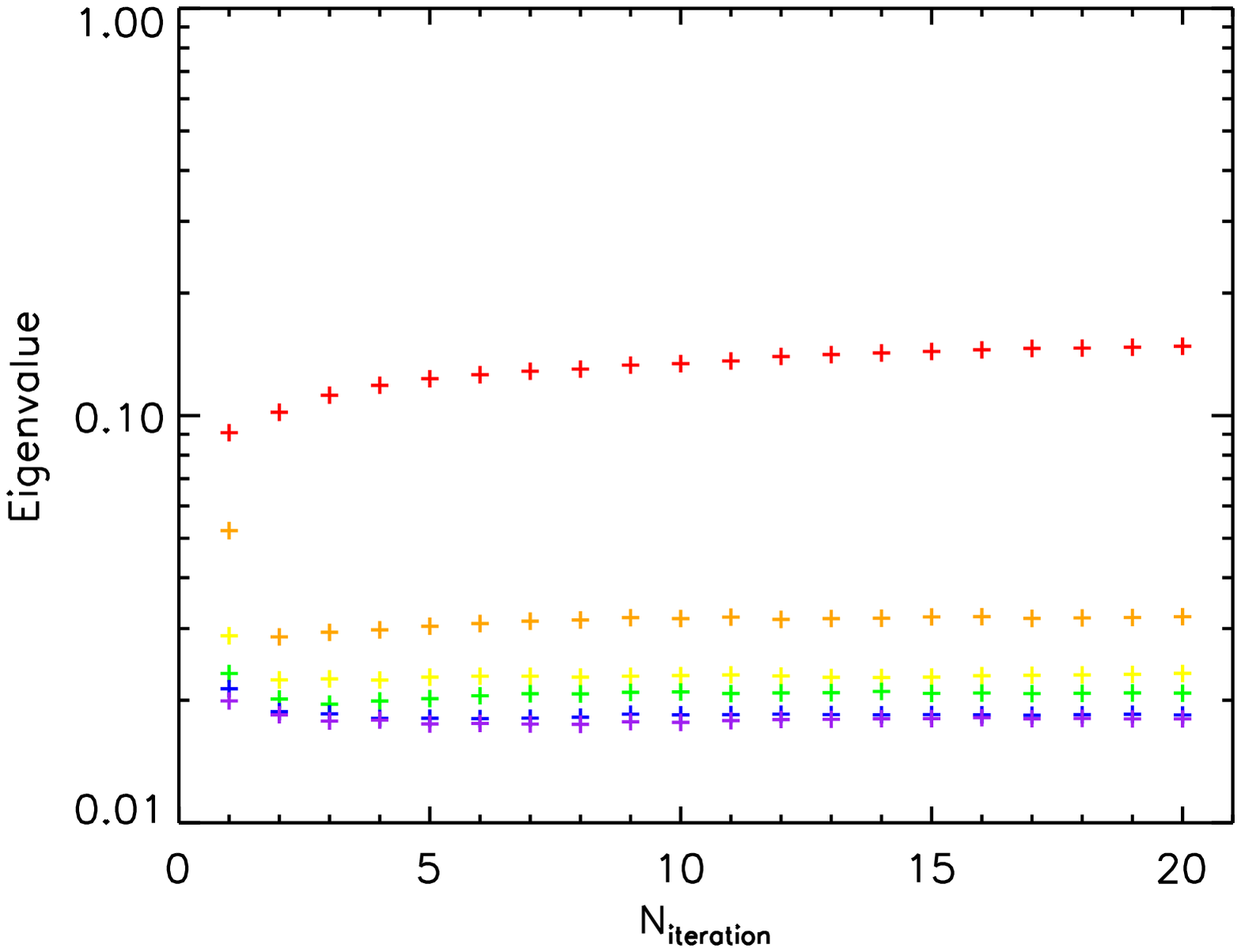}
\includegraphics[scale=0.5]{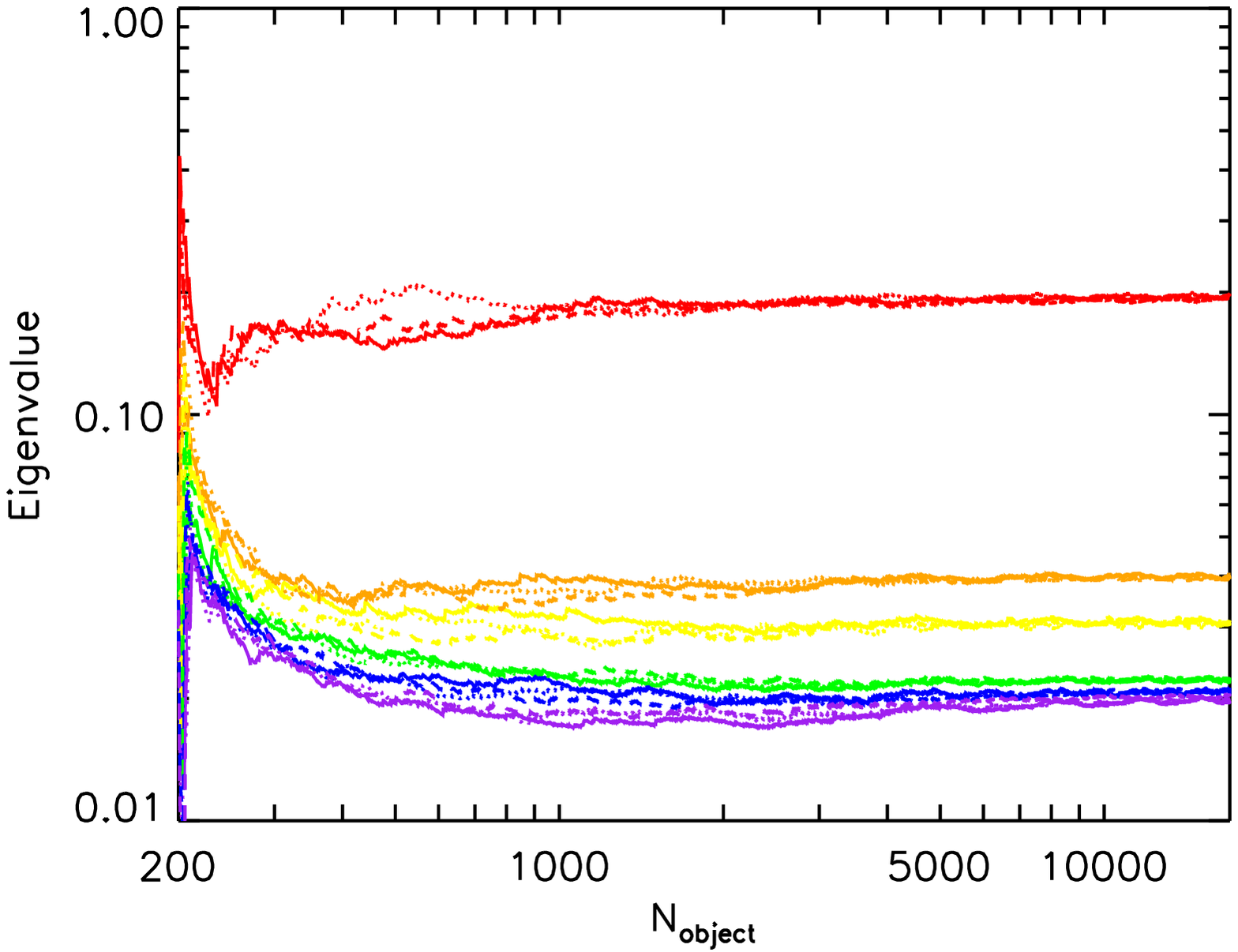}
\caption{{\it Left:} The top six normalized eigenvalues as a function of
iteration number using the classic PCA. Each eigenvalue represents the
amount of sample variance carried by the eigenspectrum. {\it Right:}
The top six normalized eigenvalues as a function of galaxy number for
robust streaming PCA shown in various colours. The x-axis begins at 200, the size of the dataset used to
initialise the eigenbasis. The dashed and dotted lines are runs from different random initialisations that
converge to the same results.}\label{fig:eval}
\end{figure*}

The VIMOS VLT Deep survey (VVDS) is a deep spectroscopic redshift
survey, targetting objects with apparent magnitudes in the range of
$17.5\le I_{AB}\le24$ \citep{vvds}. The survey is unique for high
redshift galaxy surveys in having applied no further colour cuts to
minimise contamination from stars, yielding a particularly simple
selection function, making it a very attractive dataset for
statistical studies of the high redshift galaxy population. In this
work we make use of the spectra from the publicly available first
epoch data release of the VVDS-0226-04 (VVDS-02h) field
\citep{vvds}. The spectra have a resolution of $R\!=\!227$ and a
dispersion of 7.14\AA/pixel. They have a usable observed frame
wavelength range, for our purposes, of $\sim$5500-8500\AA.

The first epoch public data release contains 8981 spectroscopically
observed objects in the VVDS-02h field, we select only those with
moderate to secure redshifts (flags 2, 3, 4, and 9) that lie in the
redshift range $0.5\!<\!z\!<\!1.0$. The redshift range is determined by the
rest-frame spectral range we have chosen for this study. The final
sample contains 3485 spectra.

The VVDS dataset provides the ideal test for a robust PCA algorithm,
because of the low signal-to-noise ratio of the spectra and the
significant chance that outliers exist due to incorrect redshift
determinations. We have chosen the 4000\AA\ break region to illustrate
the procedure because of the obvious importance of this spectral
region for galaxy evolution studies \citep[e.g.,][]{balogh99,wild07}
and also due to the wide variety of spectral features present for the
PCA to identify. Eigenspectra similar to those created in this
analysis are used by \citet{wild08} for the identification
of H$\delta$-strong galaxies in the VVDS survey.

\subsection{Classic and Trimmed Analysis}

To provide a comparison for the robust algorithm, we first perform the
classic PCA using an SVD algorithm. The spectra are corrected for
Galactic extinction assuming a uniform E(B$-$V)$=0.027$
\citep{McCracken03}, moved to the galaxy rest-frame and interpolated
onto a common wavelength grid. Regions of the spectra with bad pixels
are identified using the associated error arrays, regions
with strong night sky lines are included into the mask.  Each spectrum
is normalised, by dividing by the median flux in the good pixels, and
gaps in the dataset caused by bad pixels are filled with the median of
all other spectra at that wavelength.  The mean spectrum is calculated
and subtracted, and PCA is then performed on the residuals. The
resulting mean spectrum and eigenspectra are presented in the first column of
Figure \ref{fig:espec}.

Clearly the noise level of the eigenspectra resulting from the classic
PCA is high. The distribution of principal component amplitudes
reveals that the signal in many of the eigenspectra is dominated by a
small number of outliers. A natural way to improve on this situation
is through the iterative removal of these outliers based on the
principal component amplitude distributions.
This procedure is essentially the same as the calculation of
truncated statistics, e.g., the trimmed mean, when one excludes
some percentage of the objects symmetrically based on their rank.
For the dataset in
question, 20 iterations are required to reduce the number of $3\sigma$
outliers in the top 10 eigenspectra to half a dozen, resulting in a
total number of 2675 spectra for the final PCA. For a more thorough
analysis, a convergence criteria can be employed to indicate when the
eigenspectra cease to vary significantly \citep[e.g.,][]{yip04a}.

The resulting mean and eigenspectra from this trimmed PCA are shown in
the central column of Figure \ref{fig:espec}. As well as the
eigenspectra being visibly less noisy, the PCA now identifies more
physical features, linking together in a single eigenspectra those
features we know to be correlated, e.g., the Balmer Hydrogen line
series in the second eigenspectrum, and emission lines in the third
eigenspectrum. The left hand panel of Figure \ref{fig:eval} shows the
convergence with iteration number of the top six eigenvalues, which
represent the variance in the dataset described by each corresponding
eigenspectrum. The first eigenspectra converges quickly, after only a
few iterations. The later eigenspectra converge more slowly.

While this iterative procedure results in a clean set of eigenspectra,
the removal of outliers based on single components can easily lead to
the loss of information from the dataset. This occurs when more
unusual spectra, which would appear in later eigenspectra, are thrown
out as outliers in the top few components. Additionally, running a
full PCA for multiple iterations is undeniably an inefficient use of
computing power,
especially for large samples like the SDSS, where a single iteration
takes about 2 days.
We will now describe the application of the iterative
and robust PCA algorithm to the same noisy dataset, showing that the
same noise free and physically interesting eigenspectra are recovered,
more quickly and without the physical removal of spectra from the
dataset.

\subsection{Robust Eigenspectra}

Next we apply the streaming PCA method introduced earlier. For the
actual implementation, we utilise a Cauchy-type $\rho$-function:
\begin{equation}
\rho(t) = \frac{2}{\pi} \arctan\left(\frac{\pi}{2} \frac{t}{c^2} \right)
\end{equation}
%
%
and use the scalar $c$ to adjust the asymptotic value of the scale estimate
to match the standard deviation of a Gaussian point process.
First we perform a classic PCA on 200 randomly selected spectra to provide
the initial eigenbasis, and the initial $\sigma^2$ estimate
(eq.\ref{eqn:sig2}) is calculated from the sum of the residuals
between these 200 spectra and their PCA reconstructions. We set
$\alpha = 1-1/N$ where $N$ is the total number of galaxies in the
dataset.
We also set \mbox{$\delta=0.5$} to maximize the breakdown point,
which yields \mbox{$c\simeq{}0.787$} for our choice of $\rho$-function.
%
%
Our final results are robust to variations
in the size and content of the initialisation dataset and the precise
method used to initialise $\sigma$.
Changes to $\alpha$ and $\delta$
alter the speed of convergence and susceptibility to outliers as the
algorithm proceeds in time. Starting from the 201st spectrum, we input
the spectra into the robust streaming PCA algorithm. The right hand panel of
Figure
\ref{fig:eval} shows the progression of the eigenvalues with spectrum
number in separate colours from three alternative random initialisation shown in
solid, dotted and dashed lines. We see that full convergence of the top three eigenspectra is
reached in less than one round of the 3485 spectra;
naturally eigenspectra which carry less of the
sample variance converge more slowly as they depend on the lower order 
components, but in this example they still stabilise in less than two rounds of iterations.

The third column of Figure~\ref{fig:espec} shows the resulting mean
and top four eigenspectra. In this test case, the eigenspectra are
very similar to those from the trimmed PCA but minor improvements
are apparent. It is worth noting that the PCA algorithm is completely
independent of the order of the bins: it has no spatial coherence.
Hence the fact that our eigenspectra are smoother than the trimmed
basis is already an indication of them being more robust.

\section{Discussion} \label{sec:disc}

\begin{figure*}
\includegraphics[scale=0.7]{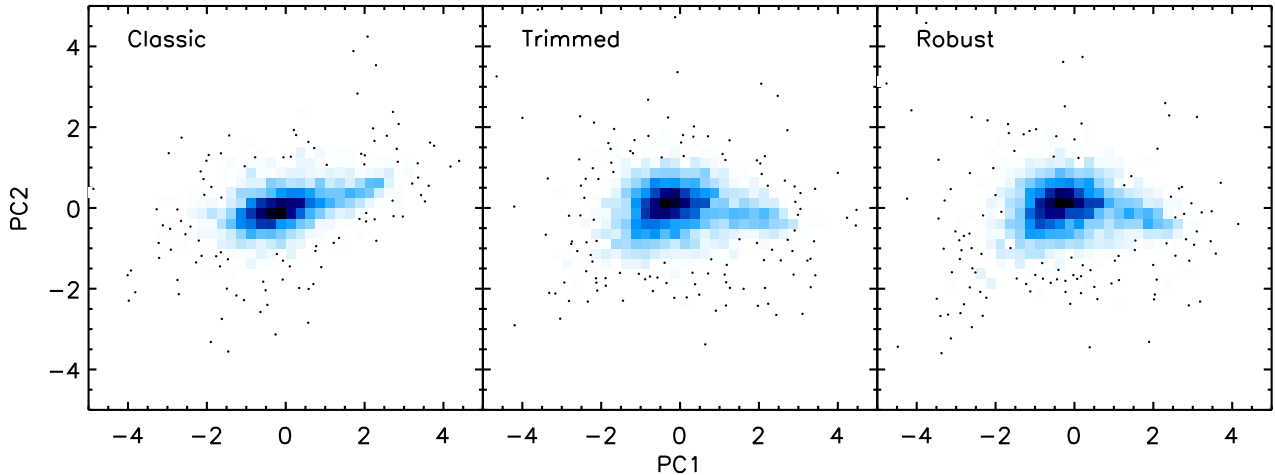}
\caption{The joint distribution of the first two principal component
amplitudes for the VVDS collection of 3485 spectra using the eigenbases presented in
Figure~\ref{fig:espec}. In order to focus on the main sample of
objects, the axes are scaled such that outliers are not
shown. From left to right: classic PCA using SVD, trimmed PCA with iterative
removal of outliers, our robust PCA from the randomised streaming algorithm.} \label{fig:pcs}
\end{figure*}

There are two important points that an effective spectral eigenbasis
must obey: (1) the eigenspectra should not introduce noise into the
decomposition of individual galaxy spectra by being noisy themselves;
(2) the top few eigenspectra must primarily describe the variance in
the majority of the dataset, and not be influenced by minority
outliers.  Additionally, the eigenbasis should be quick to calculate
and without the need for excessive memory storage.

Figure \ref{fig:espec} illustrates the success of robust statistics
for addressing point (1). The second point becomes clear when we
investigate the distribution of the principal component amplitudes of
the 3485 VVDS spectra. In Figure \ref{fig:pcs} we present the first
two principal component amplitudes for the VVDS spectra using each of
the eigenbases. The overall
correlation between these first two principal component amplitudes for
the classic PCA indicates the failure of this eigenbasis to represent
the variance in the majority of the galaxy spectra: the basis has been
influenced by outliers.

A final, important aspect of the new algorithm is the increase in
speed. The iterative truncation approach to classic PCA is clearly
inefficient, although the precise increase in speed will vary
depending on dataset properties. For our VVDS test case the 20
iterations of classic PCA take five times longer than a single iteration
of the 3485 spectra using the robust algorithm. The ratio will naturally
change in favour of the new technique for larger collections of spectra.

Our robust analysis is based on a randomised algorithm, and, as such,
it assumes that the input data entries are considered in random order both
for the initial set and the stream. When this is not the case,
the method may develop a wrong initial representation of the data,
which can take many iterations to correct.

The sensitivity of the algorithm to this issue is primarily determined
by the parameter $p$, i.e., the number of principal components kept
between steps. We expect problems only when this value is too low
compared to the weights assigned to the new input vectors. In general,
when studying an unknown dataset, we recommend that one randomises the
dataset at each iteration and solves for as many eigenspectra as
possible.

Having said that, we should also note that special ordering during the
streaming of the data might prove invaluable for studying the
evolution of the eigensystem as a function of the parameter used for
sorting the input data. However, such studies should take extreme
care in choosing the adjustable parameters (e.g., effective sample
size) and ensure that the observed trends are real and stable to
initial conditions.

\section{Summary} \label{sec:sum}

We present a novel method for performing PCA on real-life noisy and
incomplete data. Our analysis is statistically robust, and implements
the current state-of-the-art theoretical approach to generalising the
classic analysis. Our streaming technique improves the eigensystem
step-by-step when new observations are considered, and allows for
direct monitoring of the improvement. The convergence is controlled by
a single parameter that sets the effective sample size. The relevance
of this parameter becomes obvious for very large datasets such as the
SDSS catalog. These large samples are very much redundant in the
statistical sense, i.e., often the analysis of a smaller subset yields
as good results. Our method provides diagnostic tools to ensure
convergence while enabling the selection of smaller effective sample
sizes. The resulting eigenbasis has less noise than a classic PCA, and
represents the variance in the majority of the data set without being
influenced by outliers.
Compared to a common work-around for
reducing the effect of outliers on the eigenbasis by excluding extreme
instances analoguosly to the trimmed mean calculation,
the new algorithm provides a noticable improvement in robustness
and a substantial increase in speed.
A production implementation within the NVO%
\footnote{Visit the US National Virtual Observatory site at http://us-vo.org}
Spectrum Services%
\footnote{Visit the Spectrum Services at http://www.voservices.net/spectrum}
\citep{dobos}
will be released where
users of the site and Web services can direct the result sets of queries to the
robust PCA engine.
On this site we will publish the IDL scripts used for illustrations in this paper.

\section{Acknowledgments}

The authors would like to thank
the VVDS Team for making the spectra publicly available along
with their metadata. Special thanks to Bianca Garilli for her help
with obtaining the data.
TB is grateful to Ricardo Maronna for his
invaluable insights into robust statistics.
The authors acknowledge useful discussions with Gerard Lemson 
and Istv\'an Csabai.
This work was supported by
the Gordon and Betty Moore Foundation via GBMF 554 and
the MAGPop European Research Training Network, MRTN-CT-2004-503929.
AS was supported at MPA by the A.\ von Humbolt Foundation.
TB at ELTE and LD were partially supported by
MTA97-OTKA049957-NSF, NKTH:Pol\'anyi, RET14/2005, KCKHA005.

\bibliographystyle{mn2e}

\end{document}